# CHAOTIC VARIATIONS OF AES ALGORITHM

Chittaranjan Pradhan[1] and Ajay Kumar Bisoi[2]

[1]School of Computer Engineering, KIIT University, Bhubaneswar, India
`chitaprakash@gmail.com`
[2]School of Computer Engineering, KIIT University, Bhubaneswar, India
`akbisoifcs@kiit.ac.in`

## ABSTRACT

*Advanced Encryption Standard (AES) algorithm is considered as a secured algorithm. Still, some security issues lie in the S-Box and the key used. In this paper, we have tried to give focus on the security of the key used. Here, the proposed modified algorithms for the AES have been simulated and tested with different chaotic variations such as 1-D logistic chaos equation, cross chaos equation as well as combination of both. For the evaluation purpose, the CPU time has been taken as the parameter. Though the variations of AES algorithms are taking some more time as compared to the standard AES algorithm, still the variations can be taken into consideration in case of more sensitive information. As we are giving more security to the key used for AES algorithm, our proposed algorithms are very much secured from unauthorized people.*

## 1. INTRODUCTION

Due to the advancements in the Internet technology, huge digital data are transmitted over the public network. As the public network is open to all, protection of these data is a vital issue. Thus, for protecting these data from the unauthorized people, different encryption and decryption algorithms have been developed. Out of these algorithms, AES (Advanced Encryption Standard) algorithm is very much secured [1, 2].

Though AES algorithm is secured, still some security issues lie with this. In 2010, Abdulkarim Amer Shtew et. al. have found such issues and modified the standard algorithm by modifying the shift row phase involved [3]. Similarly, in 2010, El-Sayed Abdoul-Moaty ElBadawy et. al. and in 2011, Zhang Zhao et. al. have modified the standard AES algorithm by modifying the S-box generation using 1-D logistic chaos equation [4, 5]. In 2011, Alireza Jolfaei et. al. have identified such issues and modified the standard algorithm by modifying the S-box using the chaotic baker's map equations [6].

Here, we have followed a different approach by encrypting the key used in the AES algorithm. For the key encryption, we have used the 1-D logistic chaotic equation, Cross-chaotic equation and the combined version of these two techniques.

## 2. AES ALGORITHM

AES (Advanced Encryption Standard) algorithm is based on substitution and permutation principles. It takes the input text block of size 128 bit and a variable key size of 128, 192 or 256 bits for 10, 12 or 14 rounds respectively. Each round consists of several processing steps, including the encryption step itself. Similarly, set of reverse rounds are performed to transform cipher text back into plaintext [1, 2].

The pictorial representation of the AES encryption process to encrypt 128-bit PT (Plain Text) to 128-CT (Cipher Text) is shown in Figure 1. When the PT size is more than 128-bits, it will be divided into blocks of 128-bit PT. In such situation, AES encryption will be done for each block separately.

From the single key, different subkeys are generated by using the standard key schedule algorithm. So, the sensitive part of the algorithm is the secret key. Therefore we are motivated to do some processing to give further security to this key.

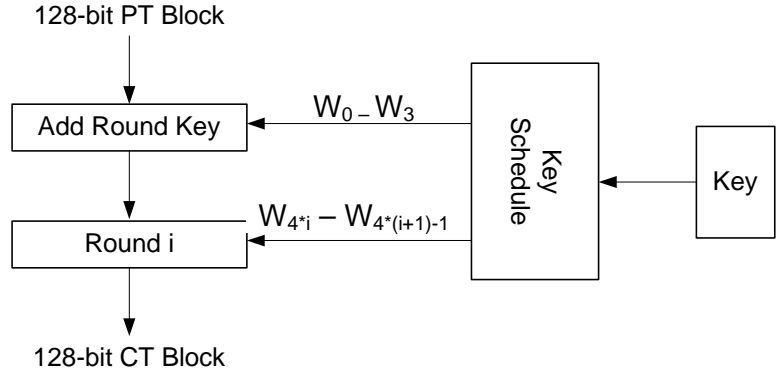

Figure 1. AES Encryption

Figure 2 shows the decryption process of the AES algorithm, which decrypts 128-bit CT(Cipher Text) to 128-bit PT(Plain Text).

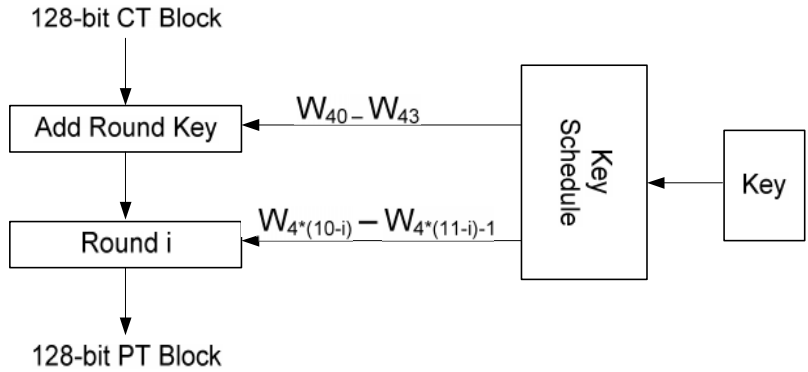

Figure 2. AES Decryption

## 3. 1-D LOGISTIC CHAOS EQUATION

Chaos signals are a kind of pseudorandom, irreversible and dynamical signals, which pose good characteristics of pseudorandom sequences. Chaotic systems are highly sensitive to initial parameters. The output sequence has good randomness, correlation, complexity and is similar to white noise. Chaotic sequence has high linear complexity and non predictability. The model here is chaos 1-D Logistic as shown in equation 1:

$$x(n + 1) = \mu \; x(n) \; [1 - x(n)] \qquad (1)$$

where, $\mu$ (0,4) and $x(n)$ (0,1). By initializing $\mu$ and $x(0)$, we can get the required chaotic signal [7]. In order to get chaotic sequences, the chaotic signal $x(n)$ must be transformed into integer sequence $s(n)$ using the quantization function as shown in equation 2:

$$f(x) = \begin{cases} 0, & 0 \le z(i,j) < 0.5 \\ 1, & 0.5 \le z(i,j) < 1 \end{cases} \quad (2)$$

where, $z(i,j)$ is the random number.

## 4. CROSS CHAOS SEQUENCE

Chaotic map is used for randomization purpose. Cross Chaotic map is defined as per equation 3:

$$\begin{cases} x_{i+1} = 1 - \mu \ y_i^2; \ x,y \ [-1,1] \\ y_{i+1} = \cos(k \ \cos^{-1} x_i) \end{cases} \quad (3)$$

where, variables $\mu$ and $k$ are the control parameters of the system. But, the system will show better chaotic behaviour when $\mu = 2$ and $k = 6$. Two chaotic sequences $X = x_0 x_1 \ldots$ and $Y = y_0 y_1 \ldots$ using initial values $x_0$, $y_0$ and control parameters $\mu$ and $k$, are generated. $X$ and $Y$ are reconstructed as row and column matrix respectively. Then they are multiplied with each other, to get a new matrix $k'$. Finally, this matrix is converted to binary matrix using the same quantization equation 2. Due to increased number of parameters, this technique gives higher security as compared to 1-D logistic map [8].

## 5. PROPOSED ALGORITHMS

Here, we have modified the standard AES algorithm by pre-processing the key using different chaotic equations. For the experimental results, we have taken lena64.bmp as the reference image which is of size 64x64. The models for different chaos variation processes are discussed below.

### 5.1. Proposed Algorithm using 1-D Logistic Chaos Equation

In this case, 1-D logistic chaos equation has been taken for encryption of the key used in AES algorithm. The block diagrams of modified AES encryption and decryption processes are shown in Figure 3 and Figure 4.

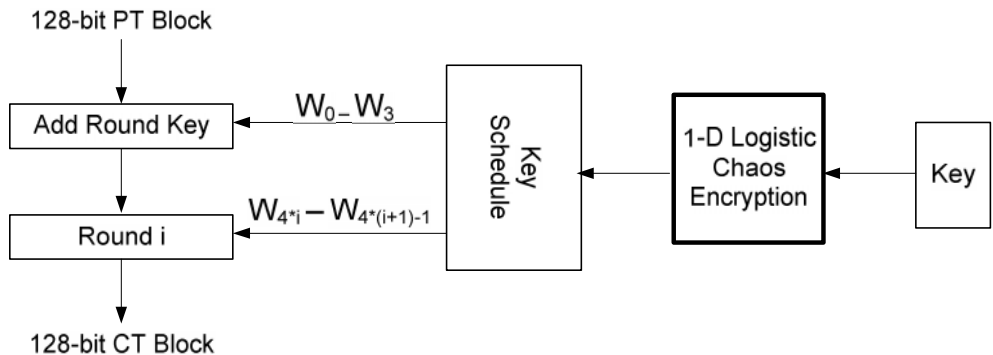

Figure 3.1-D Logistic Chaos variation AES Encryption

International Journal of Chaos, Control, Modelling and Simulation (IJCCMS) Vol.2, No.2, June 2013

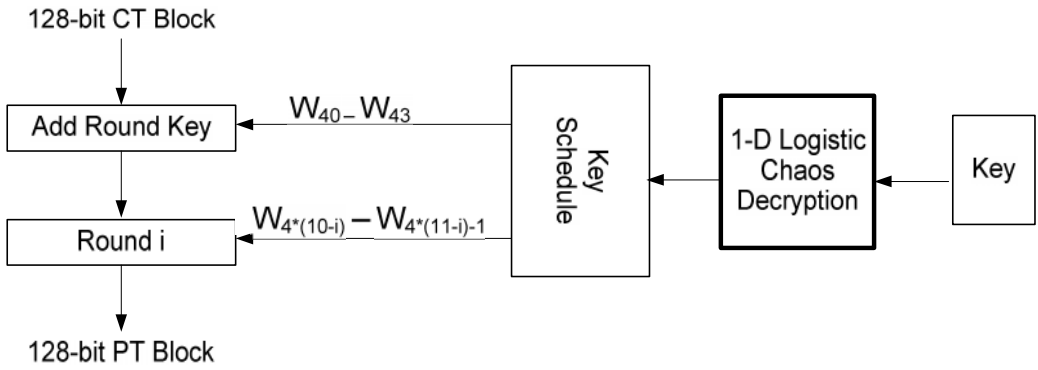

Figure 4.  1-D Logistic Chaos variation AES Decryption

Figure 5(a) shows the original image which is lena64.bmp. It has been encrypted with the modified AES algorithm using the 1-D logistic chaos encrypted key (shown in Figure 5(b)). Figure 5(c) shows the reconstructed image from the modified AES algorithm.

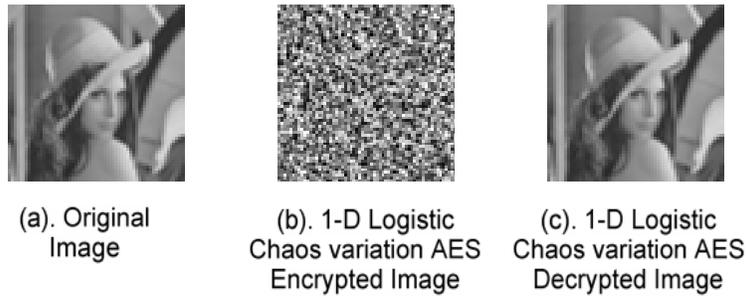

Figure 5.  1-D Logistic Chaos variation AES Process

### 5.2. Proposed Algorithm using Cross Chaos Equation

In this case, cross chaos equation has been taken for encryption of the key used in AES algorithm. The block diagrams of modified AES encryption and decryption processes are shown in Figure 6 and Figure 7.

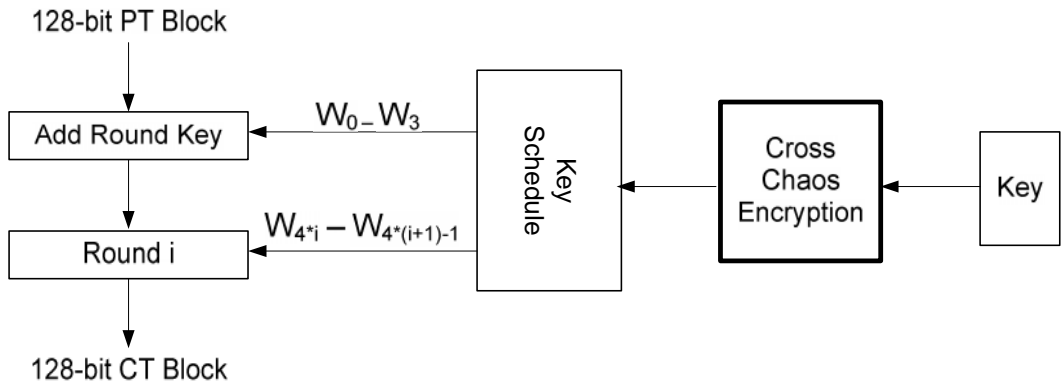

Figure 6.  Cross Chaos variation AES Encryption

International Journal of Chaos, Control, Modelling and Simulation (IJCCMS) Vol.2, No.2, June 2013

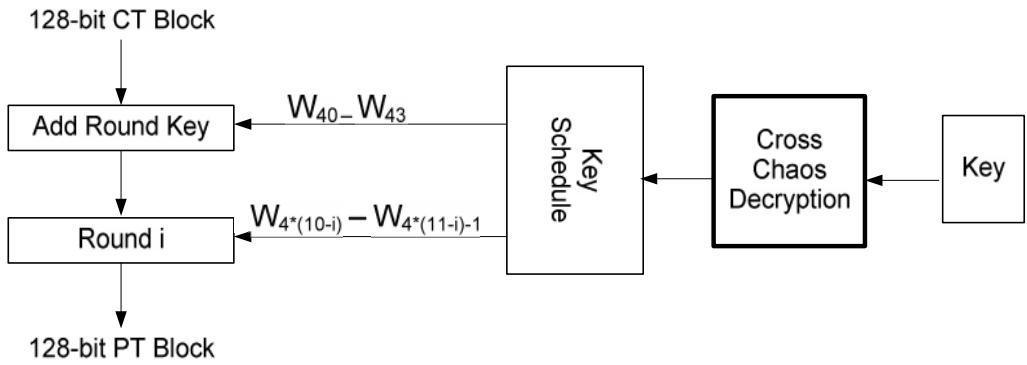

Figure 7. Cross Chaos variation AES Decryption

Figure 8(a) shows the original lena image. It has been encrypted with the modified AES algorithm using the cross chaos encrypted key (shown in Figure 8(b)). Figure 8(c) shows the reconstructed image from the modified AES algorithm.

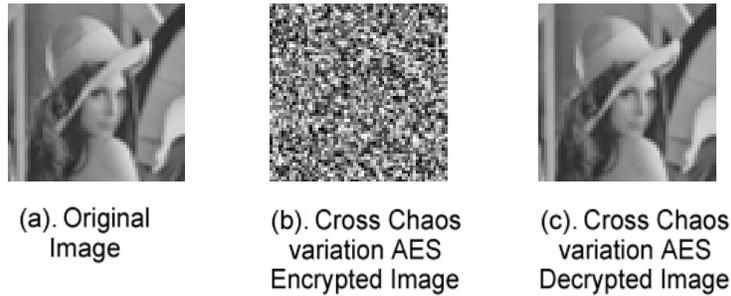

(a). Original Image  
(b). Cross Chaos variation AES Encrypted Image  
(c). Cross Chaos variation AES Decrypted Image

Figure 8. Cross Chaos variation AES Process

### 5.3. Proposed Algorithm using Dual Equation

Here, cross chaos equation along with 1-D logistic chaos equation have been taken for encryption of the key used in AES algorithm. The block diagrams of modified AES encryption and decryption processes are shown in Figure 9 and Figure 10.

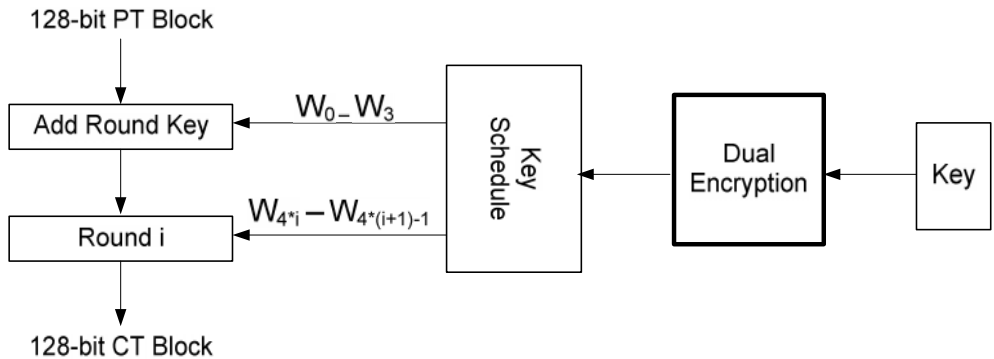

Figure 9. Dual variation AES Encryption

23

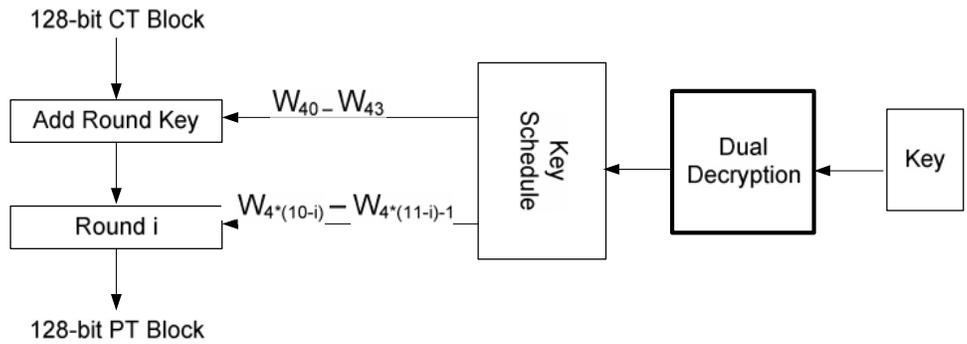

Figure 10. Dual variation AES Decryption

Figure 11(a) shows the original lena image. It has been encrypted with the modified AES algorithm using the cross chaos and 1-D logistic chaos encrypted key (shown in Figure 11(b)). Figure 11(c) shows the reconstructed image from the modified AES algorithm.

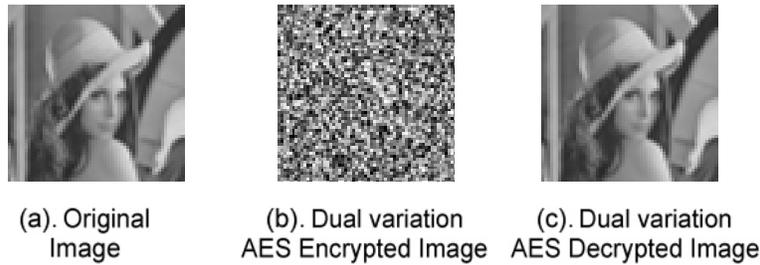

Figure 11. Dual variation AES Process

## 5.4. Analysis of Different Proposed Algorithms

We have analyzed the above discussed algorithms with the help of CPU times (in seconds). Figure 12 shows the chart for the CPU times for lena64.bmp and moon_64.bmp images in different versions of AES algorithms.

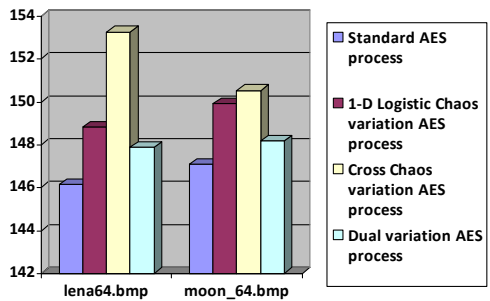

Figure 12. Chart of CPU Times

From the chart, we can find the dual variation AES algorithm though takes little more time than the standard AES algorithm, is very much secured. This is because the key has been encrypted by using cross chaos and 1-D logistic chaos equations. As different unknown parameters are involved in these two chaotic equations, it is very difficult to decrypt the key.

## 3. CONCLUSIONS

The proposed modified algorithms for the AES have been simulated and tested with different chaotic variations such as 1-D logistic chaos equation, cross chaos equation as well as combination of both. For the evaluation purpose, the CPU time has been taken as the parameter. As we are giving more security to the key used for AES algorithm, our proposed algorithms are very much secured from the unauthorized people. In future, these algorithms can be applied in some more specific fields like digital watermarking, video processing.

## REFERENCES

[1] Behrouzan A. Forouzan (2010), *Cryptography & Network Security,* TMH Publisher, ISBN: 9780070660465.

[2] Bruce Schneier (2009), *Applied Cryptography,* John Wiley & Sons Publisher, ISBN: 9780471117094

[3] Abdulkarim Amer Shtewi, Bahaa Eldin M. Hasan, Abd El Fatah .A. Hegazy (2010), "An Efficient Modified AdvancedEncryption Standard (MAES) Adapted for Image Cryptosystems", *International Journal of Computer Science and Network Security*, Vol.10, No.2, pp 226- 232.

[4] El-Sayed Abdoul-Moaty ElBadawy, Amro Mokhtar, Waleed A. El-Masry, Alaa El-Din Sayed Hafez (2010), "A New Chaos Advanced Encryption Standard (AES)Algorithm for Data Security", *International Conference on Signals and Electronic Systems,* Poland, pp 405- 408.

[5] Zhang Zhao, Sun Shiliang (2011), "Image encryption algorithm Based on Logistic chaotic system and s- box scrambling", $4^{th}$ *International Congress on Image and Signal Processing*, IEEE, pp: 177-181.

[6] Alireza Jolfaei, Abdolrasoul Mirghadri (2011), "Image Encryption Using Chaos and Block Cipher", *Computer and Information Science*, Vol. 4, No. 1, pp: 172-185.

[7] Gursharanjeet Singh Kalra, Rajneesh Talwar, Dr. Harsh Sadawarti (2011), "Robust Blind Digital Image Watermarking Using DWT and Dual Encryption Technique", $3^{rd}$ *International Conference on Computational Intelligence, Communication Systems and Networks*, IEEE, pp: 226-230.

[8] Chittaranjan Pradhan, Shibani Rath, Ajay Kumar Bisoi (2012), "Non Blind Digital Watermarking Technique Using DWT and Cross Chaos", $2^{nd}$ *International Conference on Communication, Computing & Security*, Elsevier, Vol. 6 pp: 897 – 904.

**Authors**

**Chittaranjan Pradhan** has completed his Bachelor of Engineering in Computer Science & Engineering and Master of Technology in Computer Science & Engineering. Presently, he is working as Assistant Professor in School of Computer Engineering, KIIT University, Bhubaneswar, India. He has published some innovative research papers in International Journals and Conferences. His major research areas cover Computer Security, Digital Watermarking, Image Processing, Chaos Theory and Data Mining.

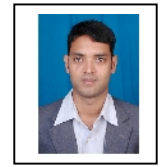

**Ajay Kumar Bisoi** obtained his Ph. D. in Particle Physics from the Utkal University, Odisha, India. Currently, he is working as Professor & Dean in School of Computer Engineering, KIIT University, Bhubaneswar, India. He is in the teaching profession since 1984. He has been a DAAD scholar at Mainz University, Germany and an Associate Member at ICTP, Italy. He has participated in many national and international forums, talks and discussions. His interest areas are: Fractal Graphics, Digital Watermarking, Chaotic Theory, Software Engineering and Computational Physics.

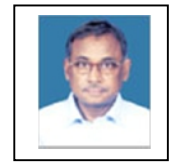